\documentclass[10pt, conference, letterpaper]{IEEEtran}
\IEEEoverridecommandlockouts      

\makeatletter
\def\ps@headings{%
\def\@oddhead{\mbox{}\scriptsize\rightmark \hfil \thepage}%
\def\@evenhead{\scriptsize\thepage \hfil \leftmark\mbox{}}%
\def\@oddfoot{}%
\def\@evenfoot{}}
\makeatother
\usepackage{cite}
\usepackage{url}
\usepackage{bm}
\usepackage{graphicx}
\usepackage{algorithm}
\usepackage[noend]{algorithmic}
\usepackage{subfig}
\usepackage{amssymb,amsmath,graphicx,charter,latexsym}
\usepackage{enumerate}
\usepackage{comment}
\usepackage{verbatim}

\newtheorem{lemma}{Lemma}

\newtheorem{theorem}{Theorem}

\allowdisplaybreaks[4]
\usepackage{bbm}

\begin{document}
\title{Index Policies for Optimal Mean-Variance Trade-Off of Inter-delivery Times in Real-Time Sensor Networks}
\author{
\IEEEauthorblockN{Rahul Singh\IEEEauthorrefmark{1},
 Xueying Guo\IEEEauthorrefmark{2} and
 P.R. Kumar\IEEEauthorrefmark{1}}

\IEEEauthorblockA{\IEEEauthorrefmark{1}Department of Electrical
and Computer Engineering, Texas A\&M University. Email: \{rsingh1,prk\}@tamu.edu}
\IEEEauthorblockA{\IEEEauthorrefmark{2}Department of Electronic Engineering,
Tsinghua University. Email: guo-xy11@mails.tsinghua.edu.cn}
\thanks{This paper is partially based on work supported by NSF under Contract Nos. CNS-1302182 and CCF-0939370, and AFOSR under Contract No. FA-9550-13-1-0008.}
 }
\maketitle
\IEEEpeerreviewmaketitle

\begin{abstract}
A problem of much current practical interest is the replacement of the wiring infrastructure connecting approximately $200$ sensor and actuator nodes in automobiles by an access point. This is motivated by the considerable savings in automobile weight, simplification of manufacturability, and future upgradability.

A key issue is how to schedule the nodes on the shared access point so as to provide regular packet delivery. In this and other similar applications, the mean of the inter-delivery times of packets, i.e., throughput, is not sufficient to guarantee service-regularity. The time-averaged variance of the inter-delivery times of packets is also an important metric.

So motivated, we consider a wireless network where an Access Point schedules real-time generated packets to nodes over a fading wireless channel. We are interested in designing simple policies which achieve optimal mean-variance tradeoff in interdelivery times of packets by minimizing the sum of time-averaged means and variances over all clients. Our goal is to explore the full range of the Pareto frontier of all weighted linear combinations of mean and variance so that one can fully exploit the design possibilities.

We transform this problem into a Markov decision process and show that the problem of choosing which node's packet to transmit in each slot can be formulated as a bandit problem. We establish that this problem is indexable and explicitly derive the Whittle indices. The resulting Index policy is optimal in certain cases. We also provide upper and lower bounds on the cost for any policy. Extensive simulations show that Index policies perform better than previously proposed policies. 
\end{abstract}
\section{Introduction}
Traditionally, throughput and delay have been used as performance metrics to judge quality of service (QoS) \cite{delay1,delay2,delay3,delay4,delay5,delay6,delay7}. The steady-state variance of inter-delivery times of packets is considered as a measure of service regularity in~\cite{atilla1}. Motivated by cyber-physical systems applications serving sensors, we address the problem of achieving an optimal ``mean-variance trade-off" in the inter-delivery times of packets of $N$ clients sharing $K$ channels.
 
We consider an access point with $K$ channels shared by $N$ clients. The clients desire a high throughput with high service regularity. We can associate a reward function $\frac{\theta_i}{\bar{D}_i} - var(D_i)$ with client $i$, where $\theta_i$ is the parameter that client $i$ uses to tune its trade-off between its throughput $\frac{1}{\bar{D}_i}$ (where $\bar{D}_i$ is the mean inter-delivery time between packets of client $i$) and the service regularity $var(D_i)$, the variance of the inter-delivery times for client $i$. By varying $\theta_i$ one can explore the full range of design freedom along the Pareto frontier of all mean-variance tradeoffs. In summary, the net function which captures the trade-off is,
\begin{align*}
\sum_{i=1}^{N} R_i\left(\frac{\theta_i}{\bar{D}_i} - var(D_i) \right),
\end{align*}
where $R_i>0$ is the weight attached to client $i$, and $\theta_i$ is a tunable parameter permitting full exploration of the Pareto frontier.

Our contributions can be summarized as follows. We show how one may obtain tractable decoupled solutions for the problem of scheduling the clients by addressing it as a Restless Multi-Armed Bandit Problem~\cite{Whi80}. In particular we obtain the Whittle indices in a closed form, which yields a very elegant solution based merely on comparing the indices of the clients. We also derive upper bounds on the achievable performance of any policy. Simulation results show that the performance of the obtained Index policy is very close to optimal.
\section{Related Works}
The steady-state variance of the inter-delivery times of packets of clients as a measure of service regularity has been considered in~\cite{atilla1}. References~\cite{atilla1} and \cite{atilla2} consider the scenario where multiple queues are sharing a server and deal with the problem of stabilizing the queues while ensuring an optimal delay and service regularity.\cite{rahul,rahul1} perform an analysis of the pathwise starvations in service for the case of a single-hop multi-user wireless network.

A detailed intoduction to Restless Multi-Armed Bandit Problems (RMBP) can be found in~\cite{Whittle2011Book}. RMBP and its relaxation were first introduced in~\cite{Whi80}. The RMBP model has been used earlier in works such as~\cite{zhao}, which considered the problem of choosing an appropriate channel for up and downlink transmissions in multichannel access. Reference~\cite{ansell} is another notable work which uses the RMBP model and derives index policies for optimizing convex holding costs in a multiclass queue.

We also note that optimality of Index policies has been established in certain cases as the population of arms goes to infinity \cite{weber} and extensive simulations have shown that Index policies have ``good" performance even in the finite population regime~\cite{ansell},\cite{Glazebrook06_machine_index}. References\cite{var1,var2,var3,var4} consider minimization of variance as an objective in Markov Decision Process.

\section{System Model}
We consider the situation where time has been discretized into slots, and the duration of a slot corresponds to the time taken to attempt a packet transmission. Each client is assumed to have one packet at the beginning of each slot. In each slot, a scheduler chooses $K$ out of the $N$ clients, and attempts to deliver their packets. Channel unreliability is modeled by supposing that if client $i$ is served in slot $t$, then the packet is delivered with probability $p_i$, independent of the past attempts. Moreover the service times are independent across clients. 
The scheduler has to choose the $K$ clients transmitted in each slot so as to maximize the reward function,
\begin{align}\label{r11}
\sum_{i=1}^{N}	R_i\left(\frac{\theta_i}{\bar{D}_i} - var(D_i)\right),
\end{align}
where $\bar{D}_i$ and $var(D_i)$ are the mean and variance of the inter-delivery times of packets for client $i$ in the steady state distribution.
\section{Markov Decision Process Formulation}
The system state at time $t$ is given by the vector $\mathbf{s}(t) := \left(s_1(t),\ldots,s_{N}(t)\right)$, with $s_i (t)$ denoting the time slots elapsed between the latest delivery of a packet of client $i$, and $t$. Because time is discretized, the state vector $\mathbf{s}(t)$ is updated only at the beginning of slot $t$, and remains unchanged within the slot. The state thus evolves as,
\begin{align*}
s_i(t+1) = 
\begin{cases}
& s_i(t) +1 \mbox{ if no packet of client } i \mbox{ is }\\
& \mbox{ delivered in slot } t,\\
& 0 \mbox{  if a packet of client } i \mbox{ is delivered in}\\ 
&\mbox{ slot } t.
\end{cases}
\end{align*} 
The Access Point (AP) takes a decision at the beginning of the slot $t$ to grant channel access to $K$ clients by choosing a control $\mathbf{u}(t)\in\{0,1\}^{K}, \sum_{i}^{N}u_i(t)=K$, where $u_i(t)=1$ implies that client $i$ will be granted channel access in slot $t$. The decision can be based on the entire past history of the system up to time $t$.
   
The ``reward earned" at time $t$ when the system is in state $\mathbf{s}$ is given by $\sum_{i=1}^{N} R_i\left( \theta_i  \mathbbm{1} \left(s_i = 0 \right) -s_i            \right) $, and thus is solely a function of the system state $\mathbf{s}$. With this set-up, the process $\mathbf{s}(t)$ becomes a controlled Markov process. 

For a positive discount factor $\beta<1$, the $\mathbf{\beta}$-discounted optimization problem is to design control policy $\mathbf{u}(t)$ so as to maximize the expected infinite horizon discounted reward,
\begin{align}\label{betamdp}
\liminf_{T\to\infty}\mathbb{E}\sum_{t=0}^{T} \beta^{t}\left(\sum_{i=1}^{N} R_i\left( \theta_i  \mathbbm{1} \left(s_i = 0 \right) -s_i            \right)\right).
\end{align}
Similarly the average reward problem is to maximize the expected infinite horizon time-average reward,
\begin{align}
\liminf_{T\to\infty}\mathbb{E}\frac{1}{T}\sum_{t=0}^{T} \left(\sum_{i=1}^{N} R_i\left( \theta_i  \mathbbm{1} \left(s_i = 0 \right) -s_i            \right)\right).
\end{align}
It is easily verified that the above reward function reduces to, 
\begin{align}
\sum_{i=1}^{N} R_i\left(\frac{\theta_i}{\mathbb{E}\left(D_i\right)} - \mathbb{E}\left(\frac{D_i \left(D_i +1\right)}{2}\right)  \right),
\end{align}
and thus differs slightly from the original reward function~(\ref{r11}).
\section{Whittle Index}
We will pose the MDP of the previous section as a Restless Multiarmed Bandit Problem (RMBP). First we briefly describe the RMBP. A detailed discussion can be found in~\cite{Whi80,Whi88,Whittle2011Book}.
 
Consider a bandit which has $N$ arms modeled as Markov processes. At each time a player can choose to play any $K<N$ arms and collect a reward from each arm, where the reward is a function of the current state of the arm that is played. The time evolution of each arm depends on whether it was chosen to play or not; thus the bandits (arms) are ``restless" and evolve even if they are not played. The player has to choose the $K$ arms to play at each time, so as to maximize the expected reward.

A  ``Whittle" policy, or ``Index-based" policy, for the RMBP, calibrates each of the $N$ arms by deriving $N$ positive functions (called ``index functions")  $W_i(\cdot), i=1,\ldots,N$, which are defined for each possible value that the state of arm $i$ can assume. At time $t$ the policy simply chooses to play the $K$ arms having the $K$ largest values of $W_i(s_i(t))$. After a re-labeling so that $W_1(s_1(t))\geq W_2(s_2(t))\geq W_N(s_N(t))$, the choices at time $t$ are
\begin{align*}
u_i(t) = 
\begin{cases}
1 \mbox{ for } i = 1,2,\ldots,K, \\
0 \mbox{ otherwise}. 
\end{cases}
\end{align*}

The derivation of the functions $W_i(\cdot)$ follows the following procedure. Each arm is considered in isolation from the rest of the arms, and the reward function is now modified so that the player receives, in addition to the original reward of the arm, a ``subsidy" each time that he chooses not to play the arm (chooses ``passive action"), and the goal once again is to maximize the average reward. After having solved this problem, let us denote by $\Pi(w)$ the set of states that an optimal policy chooses to not play arm (stay passive). Then the arm is said to be \textit{indexable} if for any two values of subsidies $w_1,w_2$, we have $w_1 >w_2 \implies \Pi(w_2)\subseteq \Pi(w_1)$, and the original MDP is said to be indexable if all the $N$ arms are indexable. 
 In case the MDP is indexable, the \textit{Whittle Index} as a function of the state value $s$ is defined as the smallest value of subsidy that makes an optimal policy choose the passive action when the client is in state $n$, i.e., 
\begin{align}\label{whittle}
W(n) = \inf \{w: n\in \Pi(w) \}.
\end{align}
Thus, the Whittle index measures, in a sense, the ``value" of an arm as a function of the present state, and the Whittle or Index policy chooses those $K$ arms which have the highest value amongst the $N$ arms.
\section{The Client Scheduling Problem is Indexable}
We will consider the $\beta$-discounted MDP, show that it is indexable and derive the corresponding Whittle index. The results for the average reward MDP will be obtained by letting $\beta\to 1$. We begin with a brief description of the single-arm $\beta$ discounted reward problem.

Consider the following single client $\beta$ discounted bandit problem parametrized by $w$ and $\beta$. The subscripts are suppressed for convenience since the discussion below applies to each of the $N$ clients. Thus $s(t),p$ are used in place of $s_i(t),p_i$.

There is a single client, whose state at time $t$, $s(t)$, is the time-elapsed-since-last-packet-delivery. At each time-slot, we can choose from the following two control actions: either attempt the transmission of a packet for it (active), or stay idle (passive). The reward earned at time $t$ is $=-Rs(t) + w+R \theta \mathbbm{1}\{s(t)=0\}$ if the client chooses the passive action of not transmitting, while a reward of $-Rs(t)+R\theta \mathbbm{1}\{s(t)=0\}$ is earned if client chooses the active action of transmitting. If the action at time $t$ is active, then $s(t+1)$, the state at time $t+1$, becomes $0$ with probability $p$, and $s(t)+1$ with probability $1-p$. If the action at time $t$ is passive, then $s(t+1) = s(t)+1$. The costs are additive over time after discounting by a factor $\beta^{t}$. A policy whether to be active or remain passive at time $t$ when the system state at time $t$ is $s(t)=s$.

We will prove that there is an optimal policy which is of \textbf{threshold type}, i.e. there is a threshold ``elapsed time since last delivery" $T$ (which depends on $\beta,w,p$), such that the policy which keeps the client passive in slot $t$ if $s(t) < T$, and active if $s(t)\geq T$, is optimal.

By $c_i(T)$ we will denote the $\beta$-discounted reward earned by a policy when the system starts with an initial state value of $i$ at time $0$, and the policy with threshold at $T$ is used. Let $\tau_i$ be the first time that state $i$ is hit, i.e. $\tau_i = \min\{t\geq 1:s(t) = i\}$. By  ``reward earned in the cycle $i\to j\to 0\to i$" we will mean the reward earned by the system starting in state $i$ in the time slots $0,\ldots,\tau_{i-1}$, while operating under the policy with threshold at $j$. 
 Expressions involving reward-functions belonging to a single value of threshold are at times not mentioned as a function of threshold. $X_p$ is a random variable that is geometrically distributed with parameter $p$. Also, we define $X:=\mathbb{E}\beta^{X_p}$ and $Y:=\mathbb{E}X_p\beta^{X_p}$.
  \begin{lemma}\label{l1_new}
Consider the single client $\beta$ discounted MDP.
	\begin{enumerate}
		\item $c_{i}(i+1)-c_{i}(i)$ is a linear increasing function of the subsidy $w$ for all $i\geq 0$ It is strictly negative when $w=0$.
		\item For each $n\geq 0$, there exists a unique value of the subsidy, denoted $W(n)$, such that $c_{n}(n+1)= c_{n}(n)$. 
		\item $W(n)\geq W(n-1)$; thus $W(n)$ form an increasing sequence.
		\item For all values of thresholds $T$, if $j>i\geq T$, then $c_i(T)>c_j(T)$.  
	\end{enumerate}
\end{lemma}

\begin{IEEEproof}
For $T\geq 0$, the infinite horizon discounted reward earned starting in state $i$ and following a policy with threshold $T+i$ is,
\begin{align*}
	c_i(i+T) &= w\sum_{j=0}^{T-1}\beta^j - \sum_{j=0}^{T-1}R\left(i+j\right)\beta^j \\
		& +R \beta^{T}\left[\mathbb{E}\left[-\sum_{j=0}^{X_p-1} \left(i+T+j\right)\beta^{j} \right]\right] \\
		& + \beta^{T}\left(\mathbb{E}\beta^{X_p}\right)\left[R\theta + \sum_{j=0}^{i}\left(w-Rj\right)\beta^{j}\right]\\
		&+\beta^{T+i}\left(\mathbb{E}\beta^{X_p}\right)c_i(i+T).
\end{align*}
Thus $c_i(i+T)$ depends on $w$ as, 
\begin{small}
	\begin{align}
		\nonumber
		& \left[{w\sum_{j=0}^{T-1}\beta^j  + w\beta^{T}\left(\mathbb{E}\beta^{X_p}\right)\sum_{j=0}^{i-1}\beta^{j}}\right]\big/\left[
		{1-\beta^{T+i}\left(\mathbb{E}\beta^{X_p}\right)}\right]\\
		\nonumber
		=& w\left[\frac{1-\beta^{T}}{1-\beta} + \beta^{T}\frac{p\beta}{p\beta + 1-\beta }\cdot \frac{1-\beta^{i}}{1-\beta}\right]\notag \big/\left(1- \beta^{T+i}\frac{p\beta}{p\beta + 1-\beta }\right) 
		\\
		\label{in4}
		=&\frac{ w \left[1-\beta+p\beta  -\beta^{T}(1-\beta +p\beta^{i+1} )\right]}{\left(1-\beta\right)\left(1-\beta+p \beta -\beta^{T+i+1}p \right)} .
	\end{align}\end{small}
Thus $c_i(i+1)-c_i(i)$ depends on $w$ as, $\frac{ w
			\left(1-\beta\right)\left(1-\beta +p\beta \right)
		}{
		\left(1-\beta+p\beta-p\beta^{i+1} \right) 
		\left( 1-\beta+p\beta-p\beta^{i+2}
		\right)}$,
which is linear and increasing in $w$.

Now we consider the case when $w=0$. If $C_1$ is the cost of cycle $i\to i \to 0 \to i$, then it follows via a simple coupling argument that the cost of cycle $i\to i+1 \to 0 \to i+1$, denoted $C_2$, is given by,
\begin{align*}
C_2 = -Ri + \beta C_1-R\beta \mathbb{E}\sum_{j=0}^{X_p-1}\beta^j,
\end{align*}
and thus to prove the second result of the first statement, we only have to show that
\begin{align*}
\frac{C_1}{1-\beta^{i}X} - \frac{-Ri + \beta C_1-R\beta \mathbb{E}\sum_{j=0}^{X_p-1}\beta^j}{1-\beta^{i}\beta X} > 0. 
\end{align*}
This is equivalent to showing that,
\begin{align*}
C_1  > -Ri\cdot \frac{1-\beta^i X}{1-\beta} -R\beta \left(\mathbb{E}\sum_{j=0}^{X_p-1}\beta^j\right) \cdot \frac{1-\beta^i X}{1-\beta}
\end{align*}
We observe that $-Ri\cdot \frac{1-\beta^i X}{1-\beta} -R\beta \left(\mathbb{E}\sum_{j=0}^{X_p-1}\beta^j\right) \cdot \frac{1-\beta^i X}{1-\beta}$ is the reward earned over the cycle $i\to i\to 0 \to i $ if one were to modify the original cost function and instead charge a penalty of  $-Ri$ for value of states $s(t)\leq i$ and a penalty of $-Rs(t)$ if $s(t)>i$. However since the original reward function is $=-Rs(t)+R\theta \mathbf{1}\{s(t)=0\}$ (note that $w=0$), a simple coupling argument shows that the reward earned is lower with the modified function. This completes the proof of first statement.

Note that from the first statement it follows that $c_n(n+1)-c_n(n)$ is a linear increasing function of $w$ which is less than $0$ at $w=0$. Hence there exists a value of $w$ such that the function $c_n(n+1)-c_n(n)$ vanishes, and moreover vanishes at a unique point since the slope of this function is strictly positive. This value of $w$, where the function $c_n(n+1)-c_n(n)$ vanishes, is $W(n)$.

Let $C_1,C_2$ be the costs of cycles $n\to n \to 0 \to n$ and $n\to n+1 \to 0 \to n$. It is seen that,
\begin{align}\label{r4}
	c_n(n) = \frac{C_1}{1-\beta^{n}X}, \quad c_n(n+1) = \frac{C_2}{1-\beta^{n}\beta X}.
\end{align}
Using a coupling argument we obtain,
\begin{align}\label{r3}
C_2 = \left(W(n)-R n\right) +\beta C_1 -R\beta \mathbb{E}\sum_{j=0}^{X_p-1}\beta^j. 
\end{align}
Combining~(\ref{r4}),(\ref{r3}) and the fact that for $w=W(n)$ we have $c_n(n) = c_n(n+1)$,
\begin{align}\label{in1}
\frac{C_1}{1-\beta^n X} &= \frac{\left(W(n)-Rn\right) +\beta C_1 -R\beta \mathbb{E}\sum_{j=0}^{X_p-1}\beta^j }{1-\beta^{n+1}X},\mbox{or ,}\notag\\
 C_1\left(1-\beta \right) &= \left(W(n)-Rn-R\beta \mathbb{E}\sum_{j=0}^{X_p-1}\beta^j\right)\left(1-\beta^n X\right).
\end{align}  
Now let us check if under the value of subsidy set to $W(n)$, we have $c_{n-1}(n) > c_{n-1}(n-1)$. If this is the case, then from the first statement of this lemma, we will deduce that $W(n-1)<W(n)$. Now,
\begin{small}
	$c_{n-1}(n) > c_{n-1}(n-1)$ is equivalent to showing
\begin{align*}
& \frac{W(n)-R(n-1) +\beta C_1 -\beta^n X\left(W(n)-R(n-1)\right)}{1-\beta^n X}  >\\
& \frac{C_1 + R\mathbb{E} \sum_{j=0}^{X_p-1}\beta^j -\beta^{n-1}X\left(W(n)-R(n-1)\right)   }{1-\beta^{n-1}X}.
\end{align*}
\end{small}
After some algebraic manipulations and using~(\ref{in1}) it can be shown that proving the above inequality is equivalent to proving $X>0$, which indeed is true. This completes the proof of third statement.

For the fourth statement, using a coupling argument, we obtain, $c_j(T) = c_i(T) - R(j-i)\sum_{j=0}^{X_p-1}\beta^j$, and hence $c_j(T)<c_i(T)$.
\end{IEEEproof}

\begin{lemma} \label{l2_new}
	Let the subsidy be $w=W(n)$. Then for the single client $\beta$ discounted MDP, 
	\begin{enumerate}
		\item $c_i(n)=c_i(n+1), \forall i\geq 0$.
		\item $ c_{i-1}(n)\geq c_{i}(n), \forall i\geq 1$.
	\end{enumerate}
\end{lemma}

\begin{IEEEproof}
Firstly recall that for subsidy $=W(n)$, $c_n(n) -c_n(n+1)=0$. Thus for $i=0,1,\ldots,n-1$,
\begin{align}\label{r5}
c_i(n) - c_i(n+1) = \beta^{n-i}\left(c_n(n) -c_n(n+1)\right) =0.
\end{align} 
For $i\geq n+1$,
\begin{align*}
c_i(n+1)-c_i(n) = \beta X\left(c_0(n+1)-c_0(n)\right) = 0,
\end{align*}
where the last equality follows from~(\ref{r5}). This proves the first statement.

To prove the second result, consider the following cases:
	\begin{enumerate}[i)]
		\item For $i> n$, Lemma~\ref{l1_new} implies that the inequality is true.
		\item For $2\leq i\leq n$,  denote $d_i$ as the cost incurred in the cycle $n\rightarrow 0 \rightarrow i-1 $.  Then both $c_i(n)$, and $c_i(n+1)$ can be derived in terms of $d_i$. When subsidy is equal to $W(n)$, we have $c_i(n)=c_i(n+1)$, i.e.,
		\begin{small}
\begin{align}
\label{eq:c1}
&-\beta^{n-i}(1-\beta)d_i =\\
& W(n)\left(\beta^n X-\beta^{n-i}  \right)+R\beta^{n-i}n-\beta^n X i \\
& +R\frac{\beta^{n-i}}{1-\beta}\left(\beta (1-X)-\beta^{i+1}X+\beta^{n+1}X^2 \right).
\end{align}\end{small}
where the first equality follows from statement $1$.
Similarly, $c_{i-1}(n)-c_{i}(n) \geq 0$ is equivalent to 
\begin{small}
\begin{align*}
&\sum_{j=0}^{n-i-1}\left(W(n)-Ri-Rj \right)\beta^j+\beta^{n-i}d_i  \geq \\
&\sum_{j=0}^{n-i}\left(W(n)-Ri+R-Rj \right)\beta^j+\beta^{n-i+1}d_i\\
&\quad -\beta^n X \left(W(n)-Ri+R \right),
\end{align*}
i.e.,
\begin{align*} 
-\beta^{n-i}(1-\beta)d_i+& R\frac{1-\beta^{n-i}}{1-\beta}+\left(W(n)-nR+R \right)\beta^{n-i} \\
&\mbox{      }\qquad -\beta^n X \left(W(n)-Ri+R \right) \geq 0\mbox{ or }
\end{align*}\end{small}
          $\frac{\left(1-\beta^n X\right)\left(\beta^i-\beta^{n+1}X\right)}{\beta^i\left(1-\beta \right)} \geq 0$,

where the second-last equivalence follows from~\eqref{eq:c1}.
We note that the last inequality holds trivially for all $\beta\in (0,1)$ and hence the statement $2$ holds for $i=2,\ldots,n$.
		
		\item $i=1$. We compare the cost incurred by the system starting in state $0$ over the cycle $0\to n\to 0$ (say $C_0$) with the cost incurred over the cycle $j\to n\to 0 \to j$ when starting in state $j$ ( denoted $C_j$) via coupling the processes associated with the two systems constructed on the same probability space. Clearly $C_0>C_j$. Thus $c_0(T)>c_j(T)$ for any value of threshold $T$.		
	\end{enumerate}
\end{IEEEproof}
\begin{lemma}\label{l3}
The function $w+ p\beta \left(c_{i}(T)-c_0(T)\right)$ ( which depends on $w,i,T$) is linear, increasing in $w$. Also,
\begin{align}\label{r12}
W(n)+ p\beta \left(c_{n+1}(n)-c_0(n)\right)=0 \mbox{ for } n = 0,1,\ldots.
\end{align}
\end{lemma}
\begin{IEEEproof}
We consider the following cases:
\begin{enumerate}[i)]
	\item For $i\leq T$, it follows from (\ref{in4}) that the function $w+p\beta \left(c_i(T)-c_0(T) \right)$ depends on $w$ as 
	\begin{align}\label{in5}
		\frac{1-\beta-p\beta+p \beta^{T-i+1}}{1-\beta+p\beta-p\beta^{T+1}}w.
	\end{align}
We have $1-\beta+p\beta-p\beta^{T+1} >0,\forall \beta <1$. Also, $1-\beta-p\beta+p \beta^{T-i+1} \geq 1-2\beta + \beta^{T-i+1} >0$ since the function
\begin{align*}
1-2\beta+\beta^k\geq 0, \forall k> 1, \beta \in (0,1).
\end{align*}
	Thus, in the expression (\ref{in5}) the coefficient of $w$ is positive. 
	\item For $i\geq T+1$, we have, 
	\begin{align*}
		c_i(T)=\mathbb{E}\sum_{j=0}^{X_p-1}\beta^j\left(-i-j \right)+ X c_0(T).
	\end{align*}
The dependence of $c_0(T)$ on $w$ can be obtained from~(\ref{in4}). Combining, $w+p\beta \left(c_i(T)-c_0(T) \right)$ depends on $w$ as, 
	\begin{align*}
		\frac{1-\beta}{1-\beta+p\beta-p\beta^{T+1}}w,
	\end{align*}
which has a positive slope with respect to $w$.
\end{enumerate}
This completes the proof of first statement.
Note that for $w=W(n)$, we have
\begin{align*}
c_n(n+1) &= c_n(n).
\end{align*}
This implies
\begin{align*}
 & -Rn+W(n) + \beta c_{n+1}(n+1) = -Rn \\
 &\qquad + \beta \left(pc_0(n) + (1-p)c_{n+1}(n)\right)\mbox{ i.e.} \\
& W(n) + \beta c_{n+1} =  \beta \left(p(c_0 + (1-p)c_{n+1}\right)\mbox{ and so }\\
& W(n) + p\beta \left(c_{n+1}-c_0\right) =0.
\end{align*}
Above, in the second implication, we have used the first statement of Lemma~\ref{l2_new} to remove the dependence of $c_i(\cdot)$ on the threshold values.
\end{IEEEproof}

\begin{theorem}
	For the $\beta$-discounted MDP with subsidy $w\in \left[ W(n),W(n+1) \right)$, the policy with threshold at $n$ is optimal. Thus the  MDP is indexable and $W(n)$ is the Whittle index when the state is $n$.
\end{theorem}
\begin{IEEEproof}
Fix a $w\in \left[ W(n),W(n+1) \right)$. If the policy is indeed optimal, then the Dynamic Programming optimality equation would be satisfied. Hence we only need to verify the inequality
\begin{align}\label{ver1}
&-Ri + w + \beta c_{i+1} \geq -Ri + \beta\left[\left(1-p\right) c_{i+1} + p c_0\right],  \notag\\
& \qquad \mbox{ for } i=0,1,\ldots,n,\notag\\
&\mbox{ or, equivalently, } \quad w+ \beta p \left(c_{i+1}-c_0\right) \geq 0, 
\end{align}
	with strict inequality holding if $w\in \left( W(n),W(n+1) \right)$, and equality holding for $i=n,w=W(n)$. Similarly for $i=n+1,n+2,\ldots$ we have to verify the inequality
	\begin{align}\label{ver2}
		\quad w+ \beta p \left(c_{i+1}-c_0\right) &\leq 0.
	\end{align} 
We will first prove~(\ref{ver1}). We use superscripts to distinguish between costs $c_i$ calculated under different values of subsidy. We have,
\begin{align*}
& w+ \beta p \left(c^{w}_{i+1}-c^{w}_0\right) \geq  W(n)+\beta p \left(c^{W(n)}_{i+1}-c^{W(n)}_0\right)\\
& = p\beta \left(c^{W(n)}_{0}-c^{W(n)}_{n+1}\right)	+ p\beta \left(c^{W(n)}_{i+1} -c^{W(n)}_{0}\right)\\
&=p\beta \left(c^{W(n)}_{i+1}-c^{W(n)}_{n+1}\right)\\
&\geq 0,
	\end{align*}
where the first inequality and equality follow from Lemma~\ref{l3}, and the last inequality follows from Lemma~\ref{l2_new}.

To prove~(\ref{ver2}) we have, 
	\begin{align*}
		w+ \beta p \left(c^w_{i+1}-c^w_0\right)& \leq  W(n+1)   +\beta p \left(c^{W(n+1)}_{i+1}-c^{W(n+1)}_0\right) \\
		& = p\beta \left(c^{W(n+1)}_{0}-c^{W(n+1)}_{n+2}\right)\\	
		&\quad + p\beta \left(c^{W(n+1)}_{i+1} -c^{W(n+1)}_{0}\right)\\
		&=p\beta \left(c^{W(n+1)}_{i+1}-c^{W(n+1)}_{n+2}\right)\\
		&\leq 0,
	\end{align*}
where first two steps follow from Lemma~\ref{l3}, and the last inequality follows from Lemma~\ref{l2_new}. This completes the optimality of the policy with threshold at $W(n)$. Following~\ref{whittle}, the Whittle index for the state $n$ is thus given by
\begin{align*}
\inf \{w: n\in \Pi(w)   \} = \inf \{w: w \geq W(n)  \} = W(n),
\end{align*}
where the first equality follows from the first statement of Theorem.
\end{IEEEproof}

We now proceed to explicitly derive the values of the indices $W(n)$.

\begin{theorem}\label{t2}
\begin{align*}
W(n) &= \frac{p\beta(f_1 - f_2 -f_3+f_4)}{f_5},\mbox{ where },\\
f_1 &= \frac{1-\beta^n}{(1-\beta)^2} \cdot \left(\left(1-X\right)\left[n(1-\beta) +\beta \right]-Y(1-\beta)\right),\\
f_2 &=\frac{\beta(1-\beta^n)-\beta^n n(1-\beta)}{(1-\beta)^2}\cdot (1-X),\\
f_3 &= \frac{1-X}{1-\beta}(1-\beta^n X),\\
f_4 &= \theta \left(1-X\right),\\
f_5 &= 1-\beta^{n}X -p\beta \left(\frac{1-\beta^{n}}{1-\beta}\right) \left(1-X\right)\\
&=\frac{1-\beta}{1-\beta+p\beta}.
\end{align*}
\end{theorem}
\begin{IEEEproof}
From~(\ref{r12}) we have,
\begin{align}\label{r6}
W(n) &= p\beta (c_0-c_{n+1}) \notag\\
&= p\beta (c_0 - c_n -\mathbb{E}\sum_{j=0}^{X_p}\beta^{j}       ).
\end{align}
Now,
\begin{align}\label{r7}
c_0-c_n = \frac{C_0 - C_n}{1-\beta^{n}\mathbb{E}\beta^{X_p}},
\end{align}
where $C_0,C_n$ are the costs over the cycles $0\to n \to 0$ and $n\to n \to 0 \to n$.
We can compute $C_0-C_n$ as,
\begin{align}\label{r8}
& C_0 - C_n = \left(\mathbb{E} \sum_{j=0}^{X_p-1} (n+j)\beta^{j}\right) (1-\beta^{n}) \\
\quad &+ \left(\sum_{j=0}^{n-1}(W(n) - j)\beta^{j}\right) (1-\mathbb{E}\beta^{X_p}) +\theta \left(1-\beta^{X_p}\right).
\end{align}
Combining~(\ref{r6},\ref{r7},\ref{r8}) and setting $\Delta = \mathbb{E} \sum_{j=0}^{X_p-1} (n+j)\beta^{j}$, we have,
\begin{small}
\begin{align*}
W(n) &= p\beta \left(	\frac{\Delta (1-\beta^{n}) + \left(\sum_{j=0}^{n-1}\left(W(n) - j\right)\beta^{j}\right) \left(1-\mathbb{E}\beta^{X_p}\right)}{1-\beta^{n}\mathbb{E}\beta^{X_p}}\right. \\
& \left. 	-\mathbb{E}\sum_{j=0}^{X_p-1}\beta^{j}  +\frac{\theta \left(1-\mathbb{E}\beta^{X_p}\right)}{1-\beta^{n}\mathbb{E}\beta^{X_p}}	\right),
\end{align*} \end{small}
or,\begin{small}
\begin{align*}
& W(n) \left[1-p\beta \cdot \frac{\left(\sum_{j=0}^{n-1}\beta^{j}\right) \left(1-\mathbb{E}\beta^{X_p}\right)}{1-\beta^{n}\mathbb{E}\beta^{X_p}}\right] = \\
& \quad p\beta \left(	\frac{\Delta (1-\beta^{n}) + \left(\sum_{j=0}^{n-1} - j\beta^{j}\right) \left(1-\mathbb{E}\beta^{X_p}\right)}{1-\beta^{n}\mathbb{E}\beta^{X_p}}	\right.\\
&\quad \left. -\mathbb{E}\sum_{j=0}^{X_p-1}\beta^{j}  +\frac{\theta \left(1-\mathbb{E}\beta^{X_p}\right)}{1-\beta^{n}\mathbb{E}\beta^{X_p}}	\right), \\
\end{align*} \end{small}
which simplifies to,
\begin{align}\label{r14}
W(n)\cdot f_5& =p\beta(f_1 - f_2 -f_3+f_4).
\end{align} 
\end{IEEEproof}

\begin{theorem}\label{mt}
The Whittle indices for the average cost MDP are given by,
\begin{align}\label{eq:blackwell}
 W^{\mbox{Avg}}(n) = \lim_{\beta \to 1} W^{\beta}(n) =  nRp\cdot \left(\frac{n}{2}+ \frac{1-p}{1+p}+\frac{1}{2}\right)+Rp\theta. 
\end{align}
\end{theorem}
\begin{IEEEproof}
The expression~\eqref{eq:blackwell} is easily derived from~\eqref{r14}. It remains to show that the quantities $ W^{\mbox{Avg}}(n) $ are indeed Whittle indices for the average-cost problem.  
Fix the subsidy to be $w$, and without loss of generality let $w\in \left(W^{\mbox{Avg}}(n),W^{\mbox{Avg}}(n+1)\right)$.
Below we use superscripts to exhibit the dependence of the cost on $\beta$. Now,
\begin{small}
\begin{align}\label{vf}
& c^\beta_0(n)  =\notag\\
&\quad  \frac{1}{1-\beta^{n} X} \cdot \left(w\frac{1-\beta^n}{1-\beta}+\frac{\beta(1-\beta^n)- n\beta^{n+1}(1-\beta)}{(1-\beta)^2}-\right. \notag\\
&\left. \beta^n \sum_{j=1}^{X_p -1}(n+j)\beta^j +\frac{R\theta}{1-\beta^{n}X} \right),\mbox{ and so }\notag\\
&\lim_{\beta\uparrow 1} (1-\beta)c^\beta_0(n) = \notag\\
& \lim_{\beta\uparrow 1}  \left(w\frac{1-\beta^n}{1-\beta^{n} X}  +\frac{\beta(1-\beta^n)- n\beta^{n+1}(1-\beta)}{(1-\beta)(1-\beta^{n} X)} \right.\notag\\
&\left.- (1-\beta) \beta^n \sum_{j=1}^{X_p -1}(n+j)\beta^j + \frac{R\theta\left(1-\beta\right)}{1-\beta^{n}X} \right) \notag\\
&=w\frac{np}{np + 1} + \frac{Rp(n^2+n)}{2(np+1)} + \frac{R p\theta }{np+1}\\
&<\infty\notag .
\end{align}\end{small}
Since for each $m$, $W^{\beta}(m) \to W^{\mbox{Avg}}(m) $, it follows from Theorem~\ref{t2} that there exists a $\beta^{\star}(w)$ such that the policy with the threshold at $n$ is optimal for the single client $\beta$-discounted MDP for all $\beta\in \left(\beta^{\star}(w),1\right)$. However since $\lim_{\beta \uparrow 1}(1-\beta) c_{0}^{\beta}(n)$ exists, the policy with threshold at $n$ is also optimal for the average cost problem.
However since $w$ can assume any value in the interval $\left(W^{\mbox{Avg}}(n),W^{\mbox{Avg}}(n+1)\right)$, the policy with threshold at $n$ is optimal for the average cost MDP for each value of subsidy $w\in \left(W^{\mbox{Avg}}(n),W^{\mbox{Avg}}(n+1)\right)$. Thus, 
\begin{align}\label{b3}
\inf    \{w: \mbox{optimal policy chooses active at } n  \}\leq W^{\mbox{Avg}}(n).
\end{align}
Similarly, picking subsidy $w<W^{\mbox{Avg}}(n)$ shows that the active action is not optimal for any value of subsidy $w<W^{\mbox{Avg}}(n)$. Hence,
\begin{align}\label{b2}
\inf\{w: \mbox{optimal policy chooses active at } n  \}= W^{\mbox{Avg}}(n),
\end{align} 
and we obtain that $ W^{\mbox{Avg}}(n)$ are indeed the Whittle indices for the average cost problem.
\end{IEEEproof}
We note that the expression~(\ref{vf}) is the average reward earned under the subsidy $w$ and threshold at $n$. We will denote this quantity as $C^{Avg}(W,n)$.

\section{Bounds on Optimal Reward.}
\begin{lemma}\label{l4}
For the average cost MDP, the reward obtained under any policy is upper-bounded by the value of the following optimization problem:
\begin{align}\label{r9}
&\max \sum_{i=1}^{N} R_i \left[\bar{D}_i^2+ \theta_i \frac{1}{\bar{D}_i} \right]\notag \\
&\mbox{ such that } \sum\limits_{i=1}^{N} \frac{1}{\bar{D}_i p_i} \leq 1, \bar{D}_i \geq 0, i =1,\ldots,N.
\end{align}
\end{lemma}

\begin{IEEEproof}
The random reward earned in time steps $1,2,\ldots,t$ is given by,
\begin{align*}
C(t) : = \sum_{i=1}^{N}  \frac{R_i}{t} \left[-\sum_{l=1}^{N_i(t)} D_i(l)^2+ \theta_i N_i(t)     \right],
\end{align*}
where $N_i(t)$ is the number of packets of client $i$ delivered by time $t$ and $D_i(l)$ is the interdelivery time of $l$-th packet of client $i$. Let us assume that the average interdelivery-time for client $i$ under a policy is equal to $\bar{D}_i$. Thus,
\begin{align}\label{in2}
\liminf_{t \to \infty} \mathbb{E}C(t) &\leq \limsup_{t \to \infty} \mathbb{E}C(t)\notag \\
& \leq  \mathbb{E}\limsup_{t \to \infty}  C(t) \notag\\
&=\mathbb{E}\limsup_{t \to \infty}  \sum_{i=1}^{N} R_i \left[\frac{\sum_{l=1}^{N_i(t)} D_i(l)^2}{t}+ \frac{\theta_i N_i(t)}{t}     \right]    \notag \\
&\leq\sum_{i=1}^{N} R_i \left[\bar{D}_i^2+ \theta_i \frac{1}{\bar{D}_i} \right]\notag,
\end{align}
where the second inequality follows from Fatou's lemma and the last is Jensen's inequality.
Thus solving the optimization problem~(\ref{r9}) gives a lower bound on the performance of any policy. We note that the constraint $\sum\limits_{i=1}^{N} \frac{1}{\bar{D}_i p_i} \leq 1, \bar{D}_i \geq 0$ is simply the capacity of the wireless channel.

\end{IEEEproof}

Next we consider the Lagrangian relaxation of the RMBP~\cite{Whi80}. For this, we relax the constraint of choosing $K$ arms at each time, to the constraint that one plays $K$ arms on average, i.e., $\lim\limits_{t\to\infty}\frac{\mbox{ Total numbers of arms played by time } t}{t} =K$. Clearly the maximum possible reward in the relaxed problem is greater than or equal to the reward earned by any policy for the original RMBP.
Also since the Index policy is the optimal solution to this relaxed problem (\cite{Whi80}), its value function serves as an upper-bound for the value function of the RMBP.
\begin{lemma}
Let $C^{Avg,i}$ be the average reward earned by the policy maximizing the single-client average reward under the subsidy $W$ (\ref{vf}). Then the reward for the average cost MDP obtained by any policy is less than or equal to,
\begin{align*}
&\inf_{W>0} \sum_{i=1}^{N}C^{Avg,i}(W) - W(N-K)\\
&=\inf_{W>0} \left(\sum_{i=1}^{N} W\frac{n_ip_i}{n_ip_i + 1} + \frac{R_ip_i(n_i^2+n_i)}{2(n_ip_i+1)} \right.\\
&\left. + \frac{R_i p_i\theta_i }{n_ip_i+1} -W\left(N-K\right)\right),\\
&=\inf_{W>0}  \left[W\left(\sum_{i=1}^{N} \frac{n_ip_i}{n_ip_i + 1}+K-N\right) + \frac{R_ip_i(n_i^2+n_i)}{2(n_ip_i+1)} \right.\\
&\left.\quad + \frac{R_i p_i\theta_i }{n_ip_i+1} \right],
\end{align*}
where $n_i$ is such that $W\in \left(W(n_i),W(n_i +1)\right)$.
\end{lemma}

\section{Optimality of Index Policy}
Now we consider several special cases of interest. 
\begin{theorem}
Consider the average cost problem for the case where all the clients are identical, i.e., $R_i \equiv 1$ and $p_i \equiv p $ for all the clients. The index policy is optimal in this case.
\end{theorem}
\begin{IEEEproof}
Firstly we note that in this symmetric case, the Index policy serves the client with the largest value of the state, i.e. the policy is, ``largest time-since-last-service-first". We will prove the result only for the case of two clients, each having channel reliability $p$. The case where there are multiple such clients follows in a straighforward manner.

Consider the time-horizon at $t$. If $\left(s_1,s_2\right)$ is the initial value of the state vector, and $R_t(\mathbf{s})$ is the maximum reward that can be earned when there are $t$ time-slots to go, then the Dynamic Programming optimality equation becomes,
\begin{align*}
 R_t\left[\left(s_1,s_2\right)\right]& =-\left(s_1 + s_2\right)+ (1-p) R_{t-1}\left[\left(s_1+1,s_2+1\right)\right] \\
&+p \max \{ R_{t-1}\left[\left(0,s_2+1\right)\right], R_{t-1}\left[\left(s_1+1,0\right)\right]\},
\end{align*}
where the optimal action corresponds to the one maximizing the expression on the right hand side. Let us assume without loss of generality that $s_1 < s_2$. Then $ R_{t-1}\left[\left(0,s_2+1\right)\right]\leq  R_{t-1}\left[\left(s_1+1,0\right)\right]$, which implies that the optimal action is to serve client $2$. 
\end{IEEEproof}
\section{Simulations}
\begin{figure}[!t]
\centering
\includegraphics[width=0.5\textwidth]{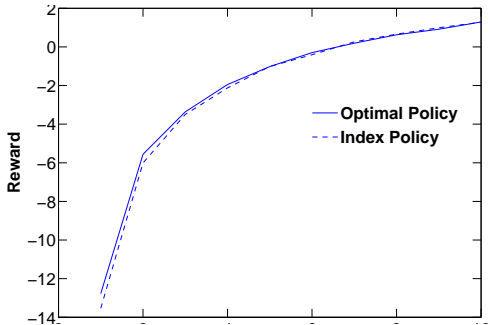}
\caption{Reward Optimal Policy vs. Index Policy for $p_1=.8,\theta_1 = 3,R_1 = 1,\theta_2=3,R_2=1$, $p_2$ varying from $.1$ to $1$.}
\label{f}
\end{figure}

\begin{figure}[!t]
	\centering
	\includegraphics[width=0.5\textwidth]{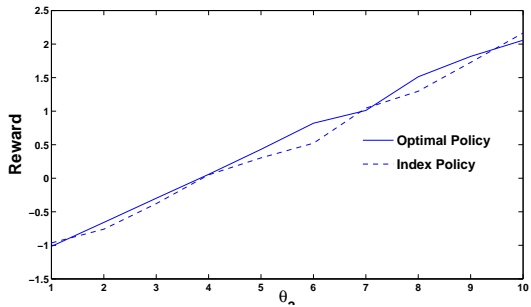}
	\caption{Reward Optimal Policy vs.Index Policy for $p_1=.8,\theta_1=3,R_1=1,p_2=.6,R_2=1$ while $\theta_2$ varies from $1$ to $10$.}
\label{g}
\end{figure}

\begin{figure}[!t]
	\centering
	\includegraphics[width=0.5\textwidth]{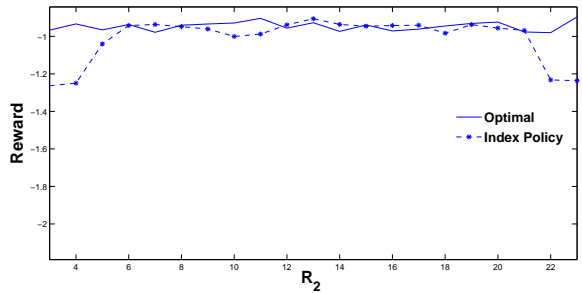}
	\caption{Reward Optimal Policy vs. Index Policy for $p_1=.8,\theta_1=5,R_1=5,p_2=.6,\theta_2=5$ while $R_2$ is varied.} 
\label{h}2
\end{figure}

We have carried out simulations to compare the performance of the optimal policy which was obtained via the Policy Iteration tool-box in Matlab vs. the Index policy which was obtained in Theorem~\ref{mt}. 
We present three plots in Figures~\ref{f}-\ref{h}. In all the cases considered $2$ clients share a single channel. 
To obtain Figure~\ref{f}, we fix client $1$'s parameter as $p_1=.8,\theta_1 = 3,R_1 = 1$, while for client $2$ we fix $\theta_2=3,R_2=1$ and vary $p_2$ from $0$ to $1$.
For Figure~\ref{g}, we fix Client $1$ parameters to be $p_1=.8,\theta_1=3,R_1=1$ while for Client $2$ we fix $p_2=.6,R_2=1$ and vary the value of $\theta_2$ from $1$ to $10$.
To obtain Figure~\ref{h}, we fix Client $1$'s parameters as $p_1=.8,\theta_1=5,R_1=5$, and for Client $2$ we fix the parameters $p_2=.6,\theta_2=5$ while varying the value of $R_2$.

We observe that Index policy gives near-optimal performance in all the cases.

\section{Concluding Remarks}
We have proposed an analytical framework for exploring the full range of  mean vs. variance tradeoffs in inter-delivery times in wireless sensor nerworks, i.e. Throughput vs. Service Regularity trade-off. The problem can be formulated as Restless Multiarmed Bandit Problem and indices can be obtained in closed form. Simulations indicate near-optimal performance of the resulting Index policy.
\bibliographystyle{IEEEtran}
\bibliography{ScheduleProblem}
\end{document}